\documentclass[fleqn,twoside]{article}
\usepackage{amsmath,amssymb}
\usepackage{epsfig}
\usepackage[headings]{espcrc2}

\readRCS
$Id: espcrc2.tex,v 1.2 2004/02/24 11:22:11 spepping Exp $
\usepackage{graphicx}
\usepackage[figuresright]{rotating}

\newcommand\Ref[1]     {Ref.\,\cite{#1}}
\newcommand\Refs[1]    {Refs.\,\cite{#1}}

\newcommand\eqns[2]    {Eqs.\,(\ref{#1}) and~(\ref{#2})}

\newcommand\fig[1]     {Fig.\,{\ref{#1}}}

\newcommand{\beq}      {\begin{equation}}
\newcommand{\eeq}      {\end{equation}}
\newcommand{\beeq}     {\begin{eqnarray}}
\newcommand{\eeeq}     {\end{eqnarray}}

\newcommand\as         {\alpha_{\rm s}}

\newcommand\rd         {{\rm d}}
\newcommand\nn         {\nonumber}

\newcommand\bom[1]     {{\mbox{\boldmath $#1$}}}

\newcommand{\bI}       {\bom{I}}
\newcommand{\bT}       {\bom{T}}

\renewcommand\O        {{\mathrm O}}
\newcommand\Oe[1]      {\ensuremath{\mathrm O(\eps^{#1})}}
\newcommand{\ep}       {\epsilon}        
\newcommand{\eps}      {\varepsilon}

\newcommand{\PS}[2]    {\rd\phi_{#1}^{#2}}
\newcommand\tsig[1]    {\sigma^{\mathrm{#1}}}
\newcommand\dsig[1]    {\rd\sigma^{{\rm #1}}}
\newcommand\dsiga[2]   {\rd\sigma^{{\rm #1,A}_{\scriptscriptstyle #2}}}

\newcommand{\rB}{{\rm B}}
\newcommand{\rR}{{\rm R}}
\newcommand{\rV}{{\rm V}}
\newcommand{\rA}{{\rm A}}
\newcommand{\rLO}{{\rm LO}}
\newcommand{\rNLO}{{\rm NLO}}
\newcommand{\rNNLO}{{\rm NNLO}}

\newcommand\ldot       {\!\cdot\!}
\newcommand\la         {\langle}
\newcommand\ra         {\rangle}

\newcommand\SME[3]     {|{\cal M}_{#1}^{(#2)}{(#3)}|^2}
\newcommand\M[2]       {\ensuremath{|{\cal{M}}_{#1}^{#2}|^2}}
\newcommand\bra[3]     {\la {\cal M}_{#1}^{#2}#3|}
\newcommand\ket[3]     {|{\cal M}_{#1}^{#2}#3\ra}
\newcommand{\mom}[1]   {\{p\}^{#1}}
\newcommand{\momt}[1]   {\{\ti{p}\}^{#1}}

\newcommand{\bA}[1]    {\bom{\mathrm A}_{#1}}
\newcommand{\bC}[1]    {\bom{\mathrm C}_{#1}}

\newcommand{\bS}[1]    {\bom{\mathrm S}_{#1}}

\newcommand{\bSCS}[1]  {\bom{\mathrm C}\kern-2pt\bom{\mathrm S}_{#1}}

\def\hP{\hat{P}}

\newcommand{\cSCS}[1]  {{\cal C}\kern-2pt{\cal S}_{#1}^{~}}

\newcommand{\ti}[1]    {\tilde{#1}}
\newcommand{\wti}[1]   {\widetilde{#1}}

\def\bsp#1\esp{\begin{split}#1\end{split}}
\def\bal#1\eal{\begin{align}#1\end{align}}

\title{Subtraction method of computing QCD jet cross sections at NNLO
accuracy}

\author{
Zolt\'an Tr\'ocs\'anyi
\address[unideb]{
University of Debrecen and Institute of Nuclear Research of the
Hungarian Academy of Sciences \\
H-4001 Debrecen P.O.Box 51, Hungary\\
E-mail: Zoltan.Trocsanyi@cern.ch},
G\'abor Somogyi
\address{
University of Z\"urich \\
Winterthurerstrasse 190, CH-8057 Z\"urich, Switzerland\\
E-mail: sgabi@physik.unizh.ch},
\thanks{
This research was supported by the Hungarian Scientific Research Fund
grant OTKA K-60432 and by the Swiss National Science Foundation (SNF)
under contract 200020-117602.}}

\runtitle{Subtraction method at NNLO}
\runauthor{Z. Tr\'ocs\'anyi}

\begin{document}

\begin{abstract}
We present a general subtraction method for computing radiative
corrections to QCD jet cross sections at next-to-next-to-leading order
accuracy. The steps needed to set up this subtraction scheme are the
same as those used in next-to-leading order computations.  However, all
steps need non-trivial modifications, which we implement such that
that those can be defined at any order in perturbation theory. We give
a status report of the implementation of the method to computing jet
cross sections in electron-positron annihilation at the
next-to-next-to-leading order accuracy.
\vspace{1pc}
\end{abstract}

% typeset front matter (including abstract)
\maketitle

\section{Introduction}

Accurate predictions of QCD jet cross sections require the computation of
radiative corrections at least at next-to-leading order (NLO) accuracy,
but in some cases also at higher order. The physical cases when
computations at the next-to-next-to-leading order (NNLO) are important
have been discussed extensively in the literature \cite{Glover:2002gz}.
Although perturbation theory is expected to be a rather systematic
procedure, this has not been reflected for many years in the computation of
radiative corrections to QCD jet cross sections. The main reason for this
is that the higher order corrections are sums of several contributions
which are separately divergent in $d = 4$ space-time
dimensions, only their sum is finite. Furthermore, these
contributions have different numbers of particles in the final state
therefore, their combination is not straightforward.

\section{Subtraction methods at NLO accuracy}

The perturbative expansion of any jet cross section can formally be written as
$
\sigma = \sigma^{\rLO} + \sigma^{\rNLO} + \sigma^{\rNNLO}
+ \ldots\;.
$
Let us consider $e^+e^- \to m \mbox{ jet}$ production, when
$\sigma^{\rLO}$ is the integral of the fully exclusive Born cross section
over the available phase space defined by the jet function $J_m$,
\beq
\sigma^{\rLO} = \int_m \rd \tsig{B}_m J_m \equiv \int \PS{m}{}\M{m}{(0)} J_m
\,.
\eeq
The NLO correction is the sum of two contributions.
We have to consider the fully exclusive cross section
$\rd\sigma^\rR$ for producing $m+1$ partons and the one-loop correction
$\rd\sigma^\rV$ to the production of $m$ partons,
\beq
\bsp
\sigma^\rNLO &= \int_{m+1}\rd\sigma^\rR J_{m+1} + \int_m\rd\sigma^\rV J_m
\\ &=
\int \PS{m+1}{}\M{m+1}{(0)} J_{m+1}
\\ &+ \int \PS{m}{}
2\mathrm{Re} \langle {\cal M}_{m}^{(1)}|{\cal M}_{m}^{(0)}\rangle J_{m}
\,.
\esp
\eeq
These two contributions are separately divergent in $d=4$ dimensions
although their sum is finite for infrared safe observables. We assume
that ultraviolet renormalization has been carried out, so the
divergences are purely of infrared origin and are regularized by
defining the integrals in $d=4-2\eps$ dimensions. 

There are several general methods of computing the finite NLO
correction. Most of these rely on the same principles, namely one
defines an approximate cross section $\rd\sigma^\rA$ that regularizes the
real correction in $d$ dimensions in all its infrared singular limits,
so the cross section 
\beq
\sigma^{\rm NLO}_{m+1} = \int_{m+1}\!\left[
  \left(\rd\sigma^\rR\right)_{\eps=0} J_{m+1}
- \left(\rd\sigma^\rA\right)_{\eps=0} J_m
\right]
\eeq
is finite.  The subtraction term in this equation is symbolic in
the sense that it is actually a sum of different terms and the jet
function depends on different momenta in each of these terms.

The first step in defining the approximate cross section is to derive
the universal factorization properties of QCD matrix elements when one
external momentum becomes soft, or collinear to another one (refered to
as unresolved). These are well-known both at NLO and at NNLO.
In writing the factorization formulae we use the
colour-state notation \cite{Catani:1996vz} and also some notation
introduced in \Ref{Somogyi:2005xz}.%
\footnote{We drop some numerical factors in order to keep the
expressions as simple as possible, as only the structure of these
formulae is relevant for the discussion.}
For the case when parton $r$ becomes soft, we have
\beq
\bsp
&\bS{r}\SME{m+1}{0}{p_r,\ldots} \propto
\\
%=-8\pi\as\mu^{2\eps} 
&\sum_{\stackrel{\scriptstyle{i,k}}{\scriptstyle{i\ne k}}}
\frac{s_{ik}}{s_{ir}s_{kr}}
\,\bra{m}{(0)}{(\ldots)} {\bom T}_{i}\ldot{\bom T}_{k} \ket{m}{(0)}{(\ldots)}
%\,,
\label{eq:SrM2}
\esp
\eeq
and for partons $i$ and $r$ becoming collinear, we have
\beq
\bsp
&\bC{ir}\SME{m+1}{0}{p_i, p_r,\ldots} \propto
\\
%= 8\pi\as\mu^{2\eps}
&\frac{1}{s_{ir}}
\bra{m}{(0)}{(p_{ir},\ldots)}
\hP_{ir}^{(0)}(z_i, k_\perp; \ep)
\ket{m}{(0)}{(p_{ir},\ldots)}
\,,
\label{eq:CirM2}
\esp
\eeq
where $s_{jl} = 2 p_j\ldot p_l$ ($j,\,l = i,\,k,\,r$), $\hP_{ir}^{(0)}$
are the Altarelli-Parisi splitting kernels and $z_i$ denotes
the momentum fraction of parton $i$ to the total momentum of the
splitting parton.  Similar formulae can be written for describing the
IR structure of the squared matrix element at NNLO. The necessary
ingredients of these factorization formulae, namely (i) the tree level
three-parton splitting functions
\cite{Gehrmann-DeRidder:1997gf,Campbell:1997hg,%
Catani:1998nv,Kosower:1999xi,DelDuca:1999ha,%
Kosower:2002su,Kosower:2003cz,Catani:1999ss}
and double soft $gg$ and $q\bar{q}$ currents
\cite{Catani:1999ss,Berends:1988zn},
(ii) the one-loop two-parton splitting functions
\cite{Bern:1994zx,Bern:1998sc,Kosower:1999rx,Bern:1999ry}
and soft-gluon current \cite{Catani:2000pi} have been known for
some time. The difficulty of using the multiple infrared factorization
formulae for constructing the approximate cross sections is amply
demonstrated by the slow progress in setting up a (general) subtraction
scheme.  

The second step is to write the IR factorization formulae in such a way
that intersecting limits can be identified and disentangled so that
multiple subtraction is avoided.  At the NLO accuracy, the only such
intersection occurs in the regions of phase space where one parton is
simultaneously soft and also collinear to a second (hard) parton and
the overlap of the soft and collinear limits can easily be identified
to be the collinear limit of the soft factorization formula
\cite{Somogyi:2005xz},
\beq
\bsp &
\bC{ir}\bS{r}\SME{m+1}{0}{p_i, p_r,\ldots}
\propto
\\ &
%= 8\pi\as\mu^{2\eps}
\frac{2}{s_{ir}}\frac{z_i}{1-z_i}\,\bT_i^2
\SME{m}{0}{p_i, \ldots}
\,.
\esp
\eeq
Thus the candidate subtraction 
\beq
\bsp
&\bA{1}\M{m+1}{(0)} =
\\
& \sum_{r}
\Bigg[\sum_{i\ne r} \frac{1}{2}\bC{ir}
+ \Bigg(\bS{r} - \sum_{i\ne r}\bC{ir}\bS{r}\Bigg)
\Bigg]\M{m+1}{(0)}
\esp
\eeq
has the same singular limit as the real correction itself and is free of
double subtractions.  However, disentangling the unresolved
limits at higher orders, when multiple soft, collinear and
soft-collinear limits overlap in a complicated way, is far more
cumbersome  \cite{Somogyi:2005xz}. This calls for a simple and
systematic procedure.  

In a physical gauge the collinear singularities are due to the
collinear splitting of an external parton
\cite{Frenkel:1976bj,Amati:1978by}.
The overall colour structure of the event does not change, the
splitting is entirely described by the Altarelli--Parisi functions
which are products of colour factors and kinematical functions
describing the dynamics of the collinear splitting. The emission of
soft gluons is just the opposite; it does not affect the kinematics of
the radiating partons, but it does affect their colour, because it always
carries away some colour charge. If we want to identify the collinear
contributions in the soft factorization formulae to {\em any order in
perturbation theory}, we can use the following simple procedure: (i)
employ the soft insertion rules \cite{Bassetto:1984ik,Catani:1999ss} to
obtain the usual expression
\beq
\bsp &
\bS{r} \SME{m+1}{0}{p_r,\dots} \propto
\sum_{i=1}^m \sum_{k=1}^m \sum_{\rm hel.}
\eps_\mu(p_r) \eps_\nu^*(p_r)
\\ &
\times \frac{2 p_i^\mu p_k^\nu}{s_{ir} s_{kr}}
\bra{m}{(0)}{(\dots)}\bT_i\ldot \bT_k\ket{m}{(0)}{(\dots)}
\,,
\esp
\eeq
with 
\beq
\quad
 \sum_{\rm hel.} \eps_\mu(p_r) \eps_\nu^*(p_r) =
-g^{\mu\nu} + \frac{p_r^\mu n^\nu + p_r^\nu n^\mu}{p_r\ldot n}
\,;
\eeq
(ii) fix the gauge vector to $n^\mu = Q^\mu - p_r^\mu\,Q^2/s_{rQ}$,
$s_{rQ} = 2 p_r\ldot Q$ ($Q^\mu$ is the total incoming momentum)
to identify the collinear contribution in the colour-diagonal terms 
\beq
\bsp &
\bS{r} \SME{m+1}{0}{p_r,\dots} \propto
\\&
\sum_{i=1}^m \Bigg[
  \frac12 \sum_{k\ne i}^m
\Bigg(\frac{2 s_{ik}}{s_{ir} s_{rk}}
- \frac{2 s_{iQ}}{s_{rQ} s_{ir}}
- \frac{2 s_{kQ}}{s_{rQ} s_{kr}}\Bigg)
\\&\qquad\qquad\quad\times
\bra{m}{(0)}{(\dots)}\bT_i\ldot \bT_k\ket{m}{(0)}{(\dots)}
\nn
\esp
\eeq
\vspace*{-1ex}
\beq
\qquad
- \bT_i^2 \frac{2}{s_{ir}} \frac{s_{iQ}}{s_{rQ}}
\SME{m}{0}{\dots}
\Bigg]
\,;
\eeq
(iii) define momentum fractions in the Sudakov parametrization of momenta
$p_i^\mu$ and $p_r^\mu$ being collinear as
$z_i = \frac{s_{iQ}}{s_{iQ}+s_{rQ}}$, so that the colour-diagonal terms
become equal to the collinear limit of the soft factorization formula.
Then the pure soft contributions are given by
\beq
\bS{r}^{\rm pure} \SME{m+1}{0}{p_r,\dots} \propto
\nn
\eeq
\beq
\bsp &
\sum_{i=1}^m \Bigg[
  \frac12 \sum_{k\ne i}^m
\Bigg(\frac{2 s_{ik}}{s_{ir} s_{rk}}
- \frac{2 s_{iQ}}{s_{rQ} s_{ir}}
- \frac{2 s_{kQ}}{s_{rQ} s_{kr}}\Bigg)
\\ &\qquad\qquad\qquad\times
\bra{m}{(0)}{(\dots)}\bT_i\ldot \bT_k\ket{m}{(0)}{(\dots)}
\Bigg]
\,.
\esp
\eeq
We checked explicitly that this procedure leads to non-overlapping
factorization formulae that describe the analytic behaviour of the
squared matrix elements in any IR limit at the NNLO accuracy
\cite{Nagy:2007mn}.

The third step is the definition of the subtraction terms. The
factorization formulae in \eqns{eq:SrM2}{eq:CirM2} (and similar ones at
NNLO) are valid in the strict limits only and have to be extended over
the whole phase space. This extension requires momentum mappings
$\mom{}_{m+1} \rightarrow \momt{}_m$ that
\begin{itemize}
\itemsep=-2pt
\item implement exact momentum conservation,
\item lead to exact phase-space factorization,
\item can be generalized to any number of unresolved partons and
\item respect the (delicate) structure of cancellations among the
subtraction terms.
\end{itemize}
For any such mapping we can write the exact factorization of the phase
space in the following symbolic form
\beq
\PS{m+1}{}(p_1,\dots;Q) =
\PS{m}{}(\wti{p}_1,\dots;Q) [\rd p_r]\,.
\eeq
Then the singular integral over the momentum of the unresolved
parton $r$ can be computed independently of the jet function and
the rest of the phase-space integration, leading to
\beq
\int_1 \rd\sigma^\rA = \rd\sigma_m^\rB \otimes \bI(\ep)
\,,
\label{eq:intsigmaA}
\eeq
where $\bI(\ep)$ is an operator in colour space with universal pole
part,
\beq
\bI(\ep) \propto \frac{\as}{2\pi}
\sum_i \left[
  \frac{1}{\ep} \gamma_i
- \frac{1}{\ep^2} \sum_{k\ne i} \bT_i\cdot\bT_k
\left(\frac{4 \pi \mu^2}{s_{ik}}\right)^\ep\right]
\nn
\eeq
\vspace*{-1ex}
\beq
+ \Oe{0}
\,,
\eeq
with $\gamma_i$ being the flavour constants defined in for instance,
\Ref{Catani:1996vz}.

The fourth step is to identify the universal IR pole structure of
one-loop QCD matrix elements,
\beq
\ket{m}{(1)}{(\mom{})} = -\frac12
\bI(\ep)\ket{m}{(0)}{(\mom{})}
+ \Oe{0}\,.
\eeq
We observe that the poles in the insertion operator $\bI(\ep)$ are
equal, but opposite in sign to the poles of the virtual correction, so
that the $m$-parton integral
\beq
\sigma^{\rNLO}_m = 
\int_m\!\left[\rd\sigma^\rV + \int_1 \rd\sigma^\rA \right]_{\eps=0}
\label{eq:rNLOm}
\eeq
is finite and we can take the physical limit $\ep \to 0$. Therefore,
the sum of the two finite contributions $\sigma^{\rm NLO}_m$ and
$\sigma^{\rm NLO}_{m+1}$ can be computed numerically
in four dimensions and is equal to $\sigma^{\rNLO}$.  

\section{A subtraction scheme at NNLO accuracy}

The physical motivation for higher accuracy and the success of the
subtraction schemes at NLO lead one to consider the extension of the
subtraction method to NNLO, when three terms contribute: the
double-real, the real-virtual and the double-virtual cross sections,
\beq
\bsp &
\sigma^{\mathrm{NNLO}} =
\tsig{RR}_{m+2} + \tsig{RV}_{m+1} + \tsig{VV}_{m}
\\ &
 \equiv
  \int_{m+2} \dsig{RR}_{m+2} J_{m+2}
+ \int_{m+1} \dsig{RV}_{m+1} J_{m+1}
\\ &
+ \int_{m} \dsig{VV}_{m} J_{m}
\,.
\esp
\eeq
The reorganization of the NNLO contributions into three finite cross
sections,
\beq
\sigma^{\rNNLO} = \tsig{\rNNLO}_{m+2} + \tsig{\rNNLO}_{m+1} +
\tsig{\rNNLO}_{m}
\,,
\eeq
is governed by the jet function as follows:
\beq
\bsp &
\tsig{\rNNLO}_{m+2} =
 \int_{m+2} \Big\{ \dsig{RR}_{m+2} J_{m+2}
-\dsiga{RR}{2}_{m+2} J_m
\\ &\qquad
-\Big( {\dsiga{RR}{1}_{m+2}} J_{m+1}
-\dsiga{RR}{12}_{m+2} J_m\Big)\Big\}
\,,
\label{eq:rNNLOm+2}
\esp
\eeq
\beq
\bsp &
\tsig{\rNNLO}_{m+1} =
\! \int_{m+1}\!\!\!\Big\{\Big( \dsig{RV}_{m+1}
+\int_1\!\!\dsiga{RR}{1}_{m+2}\Big) J_{m+1}
\\ &\qquad
-\Big[\dsiga{RV}{1}_{m+1}
+\Big(\int_1 \dsiga{RR}{1}_{m+2}\Big){}^{\mathrm{A}_1}\Big] J_{m} \Big\}
\label{eq:rNNLOm+1}
\,,
\esp
\eeq
%and
\beq
\bsp &
\tsig{\rNNLO}_{m} =
 \int_m \Big\{ \dsig{VV}_{m}
+\int_2\Big( \dsiga{RR}{2}_{m+2}
-\dsiga{RR}{12}_{m+2}\Big)
\\ &\qquad
+\int_1\Big[\dsiga{RV}{1}_{m+1}
+\Big(\int_1 \dsiga{RR}{1}_{m+2}\Big){}^{\mathrm{A}_1}\Big]\Big\} J_m
\,.
\label{eq:rNNLOm}
\esp
\eeq
Here we see that at NNLO accuracy one has to disentangle the overlapping
singularities also among the singly- and doubly-unresolved limits.
The purpose of the approximate cross section $\dsiga{RR}{12}_{m+2}$
is to regularize the singly-unresolved limits of $\dsiga{RR}{2}_{m+2}$
{\em and} the doubly-unresolved limits of $\dsiga{RR}{1}_{m+2}$
{\em simultaneously}. Similarly, the approximate cross section 
$\Big(\int_1 \dsiga{RR}{1}_{m+2}\Big){}^{\mathrm{A}_1}$ regularizes the
singly-unresolved limits of $\int_1 \dsiga{RR}{1}_{m+2}$ and the $\ep$
poles of $\dsiga{RV}{1}_{m+1}$, respectively.  This puts severe constraints
on the phase-space mappings needed for the definition of the subtractions. 

In \Ref{Somogyi:2006db} we introduced new types of momentum mappings, one
for collinear- and another for soft-type subtractions, that can easily
be generalized to any order in perturbation theory. The key feature of
these mappings is that in the factorized $m$-particle phase space all
momenta take away the recoil instead of a single one as in the case of
the dipole (or antennae) subtractions. In this way the factorization of
the phase space can be done in a way that respects the (delicate)
structure of cancellations among the various subtraction terms.

\begin{figure*}
~\quad
\includegraphics[scale=0.36]{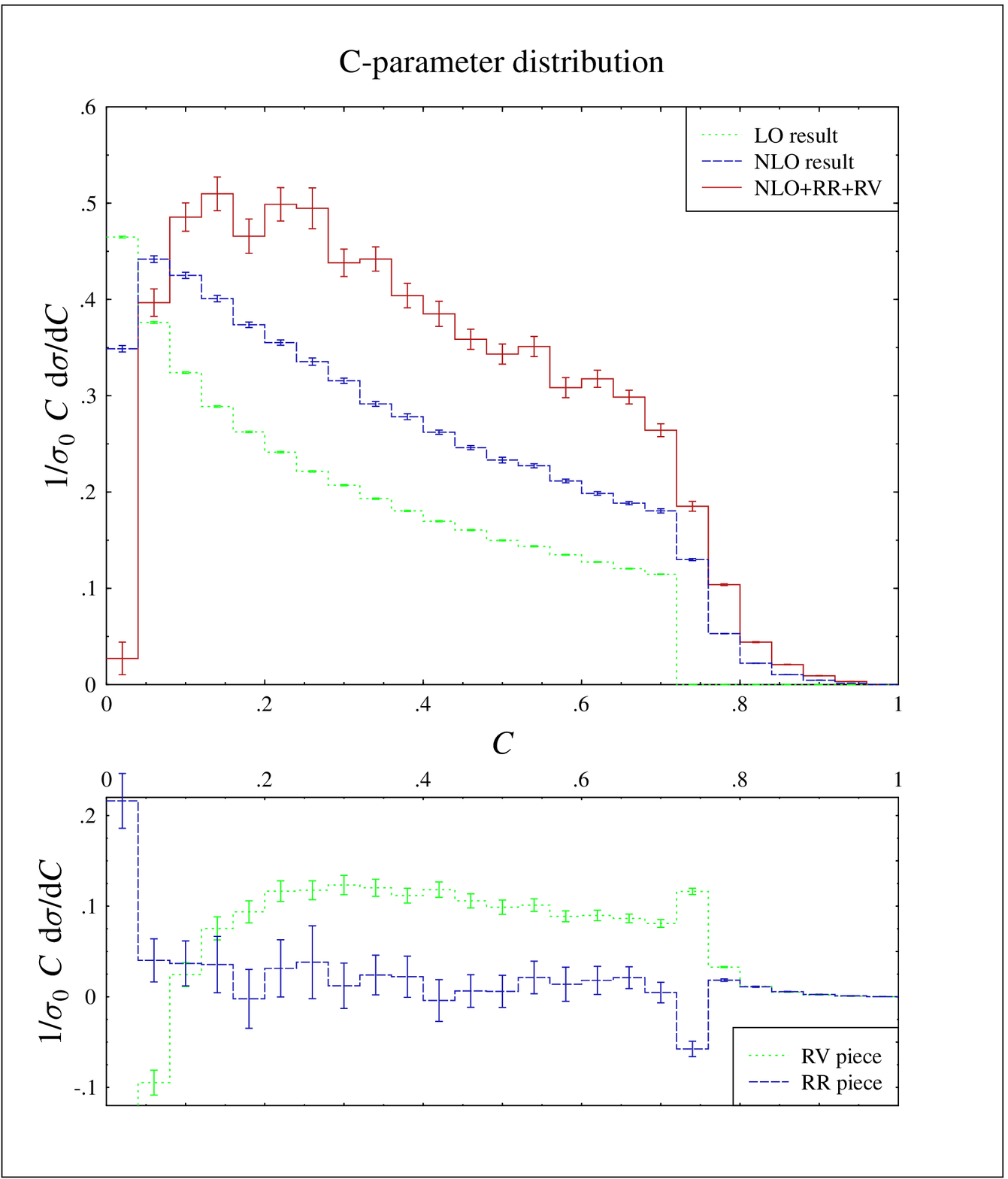}
\hfill
\includegraphics[scale=0.36]{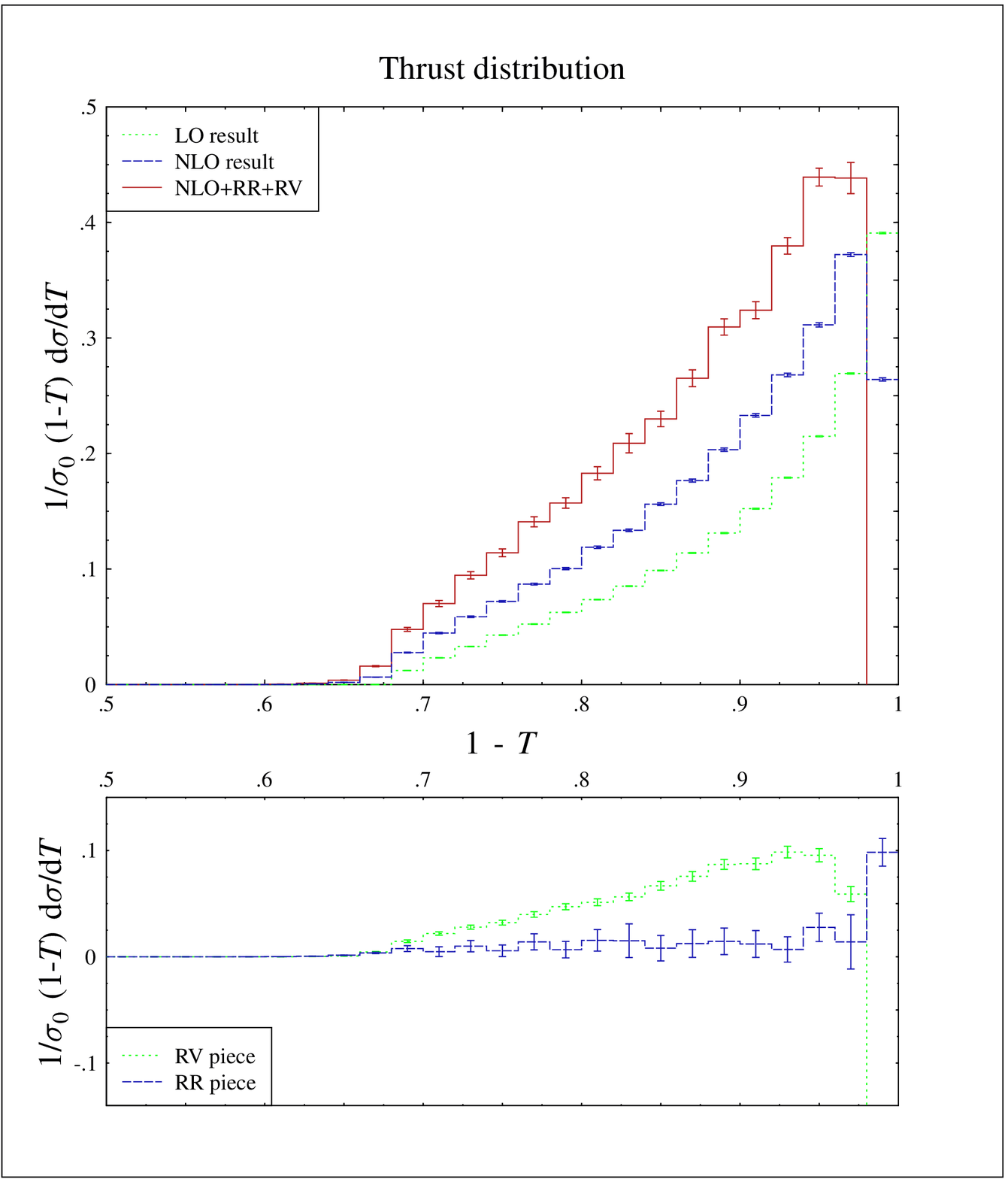}
\quad~
\caption{The five- and four-parton contributions (lower panels) to the
C-parameter (left) and thrust (right) distributions. The upper panels
show the distributions at the LO, NLO and incomplete (without the
three-parton contribution) NNLO accuracy.}
\label{fig}
\end{figure*}

The complete subtraction scheme at NNLO, based on these new, {\em fully
local} approximate cross sections
is defined in \Refs{Somogyi:2006da,Somogyi:2006db,Somogyi:2006cz}.
We employed this subtraction scheme for computing the finite cross sections
$\tsig{\rNNLO}_{m+2}$ and $\tsig{\rNNLO}_{m+1}$ of the C-parameter and
thrust distributions in electron-positron annihilation. \fig{fig} shows
the distributions normalised to the total cross section at $\O(\as^3)$
accuracy. The computer time needed for obtaining these distributions is
fairly little. The plots shown here can be
obtained on a desktop computer in about 50 hours. The still missing
three-parton contribution is a smooth function as compared to the four-
and five-parton contributions, therefore its numerical integration does
not raise any serious stability issues.

In order to have the complete physical prediction we also have to add
$\tsig{\rNNLO}_m$, which requires the integration of the subtraction
terms over the singly- and doubly-unresolved factorized phase spaces.

The necessary one-particle integrals have been computed in
\Refs{Somogyi:2008??,Aglietti:2008??,Bolzoni:2008??}.  In
\Ref{Somogyi:2008??} we used standard techniques of partial
fractioning, iterated sector decomposition
\cite{Heinrich:2008si,Bogner:2007cr} and residuum subtraction to find
the Laurent expansion of the one-particle integrals in
\beq
\int_1\Big[\dsiga{RV}{1}_{m+1}
+\Big(\int_1 \dsiga{RR}{1}_{m+2}\Big){}^{\mathrm{A}_1}\Big]
\,.
\eeq
With this technique the expansion coefficients are given as finite
integrals. In order to find explicit analytic expressions, one can use
the integration-by-parts technique to reduce the integrations to
master integrals to be evaluated by solving differential equations
\cite{Aglietti:2008??}. An alternative solution is to employ 
the Mellin-Barnes technique to evaluate the expansion coefficients in
terms of harmonic sums \cite{Bolzoni:2008??}. Our study shows that the
latter technique is in general more efficient. We expect that the same
techniques can also be employed for the computation of the coefficients
in the $\ep$-expansion of the two-particle integral
\beq
\int_2\Big( \dsiga{RR}{2}_{m+2} -\dsiga{RR}{12}_{m+2}\Big)
\,.
\eeq
This work is in progress.

In this contribution we have discussed how the same steps of setting up a
subtraction scheme for computing QCD jet cross sections at NLO accuracy
can be used, but need to be modified when generalizing to NNLO accuracy.
We introduced these modifications such that a systematic perturbative
expansion can in principle be given at any order of perturbation theory.
The explicit computations of course become rather cumbersome already at
NNLO.

We are grateful to our collaborators: U.~Aglietti, P.~Bolzoni, V.~Del
Duca, C.~Duhr and S.~Moch.

\end{document}